\documentclass[sigconf]{acmart}

\AtBeginDocument{%
  \providecommand\BibTeX{{%
    \normalfont B\kern-0.5em{\scshape i\kern-0.25em b}\kern-0.8em\TeX}}}

\usepackage{graphicx}
\usepackage{lipsum}

\newcommand\blfootnote[1]{%
  \begingroup
  \renewcommand\thefootnote{}\footnote{#1}%
  \addtocounter{footnote}{-1}%
  \endgroup
}

\begin{document}

\title{Tripartite Heterogeneous Graph Propagation for Large-scale Social Recommendation}
\author{Kyung-Min Kim$^{1,*}$, Donghyun Kwak$^{2,*}$, Hanock Kwak$^{3,*}$, Young-Jin Park$^{4,*}$}
\author{Sangkwon Sim$^{1}$, Jae-Han Cho$^{1}$, Minkyu Kim$^{1}$, Jihun Kwon$^{4}$, Nako Sung$^{1}$, Jung-Woo Ha$^{1}$}
\email{{kyungmin.kim.ml, donghyun.kwak}@navercorp.com}
\email{hanock.kwak2@linecorp.com}
\email{{young.j.park, jaehan.cho, min.kyu.kim, jihun.kwon, andy.sangkwon, nako.sung, jungwoo.ha}@navercorp.com}
\affiliation{%
  \institution{$^{1}$Clova AI Research, NAVER Corp., $^{2}$Search Solution Inc., $^{3}$LINE Plus Corp. and $^{4}$Naver R\&D Center, NAVER Corp.}
  \city{Seongnam}
  \state{Gyeonggi}
  \country{South Korea}
  \postcode{13561}
}

\renewcommand{\shortauthors}{Kim, Kwak, and Park et al.}

\begin{abstract}
\blfootnote{$^*$Authors contributed equally to this research. The authors are sorted by alphabetical order.}
  Graph Neural Networks (GNNs) have been emerging as a promising method for relational representation including recommender systems.
  However, various challenging issues of social graphs hinder the practical usage of GNNs for social recommendation, such as their complex noisy connections and high heterogeneity. The oversmoothing of GNNs is a obstacle of GNN-based social recommendation as well.
  Here we propose a new graph embedding method Heterogeneous Graph Propagation (HGP) to tackle these issues.
  HGP uses a group-user-item tripartite graph as input to reduce the number of edges and the complexity of paths in a social graph.
  To solve the oversmoothing issue, HGP embeds nodes under a personalized PageRank based propagation scheme, separately for group-user graph and user-item graph.
  Node embeddings from each graph are integrated using an attention mechanism.
  We evaluate our HGP on a large-scale real-world dataset consisting of 1,645,279 nodes and 4,711,208 edges.
  The experimental results show that HGP outperforms several baselines in terms of AUC and F1-score metrics. 
  
\end{abstract}



\maketitle

\section{Introduction}
Graph Neural Networks (GNNs) \cite{kipf2017semi,hamilton2017inductive} have gained remarkable attention with their ability to learn representations from graph data. 
The GNNs can competitively exploit graph structures via techniques such as neighborhood aggregation and pooling \cite{peter2018relational}.
So far, many variants of GNNs with different aggregation and pooling schemes have been proposed, and they are achieving promising performances in diverse fields including semi-supervised classification \cite{kipf2017semi}, drug discovery \cite{gilmer2017neural} and knowledge-based question answering \cite{sorokin2018modeling}.

Recently, deep learning-based methods have shown promising results in recommender systems \cite{haldar2018applying,wang2018billion,rohde2018recogym}.
The GNNs are becoming increasingly popular methods to leverage relational information, successfully applied in IT industries, e.g., Pinterest \cite{ying2018graph} and Alibaba \cite{cen2019representation}.
They represent user-item interactions as user-item graph and compute the similarity between nodes to recommend items to a user.
Besides, it is beneficial to use additional user-user interaction information in social recommender systems to alleviate cold-spots in a user-item graph \cite{fan2019graph}. 
However, it has not been straightforward to apply GNNs to social recommendation tasks due to the following challenging issues: 
1) As the number of user node increases, the number of edges in the user-user graph grows exponentially in general.
2) Tractable social recommendation using GNNs requires proper computational tricks and sampling methods to handle large-scale social graphs.
3) Previous GNNs suffer from the oversmoothing problem, which is to ignore the local structure as the number of layers increases \cite{xu2018representation}.
4) There are two inherently different graphs, i.e., user-user graph and user-item graph. A model has to combine these two graphs coherently.

\begin{figure}[!t]
  \centering
  \includegraphics[width=8.5cm]{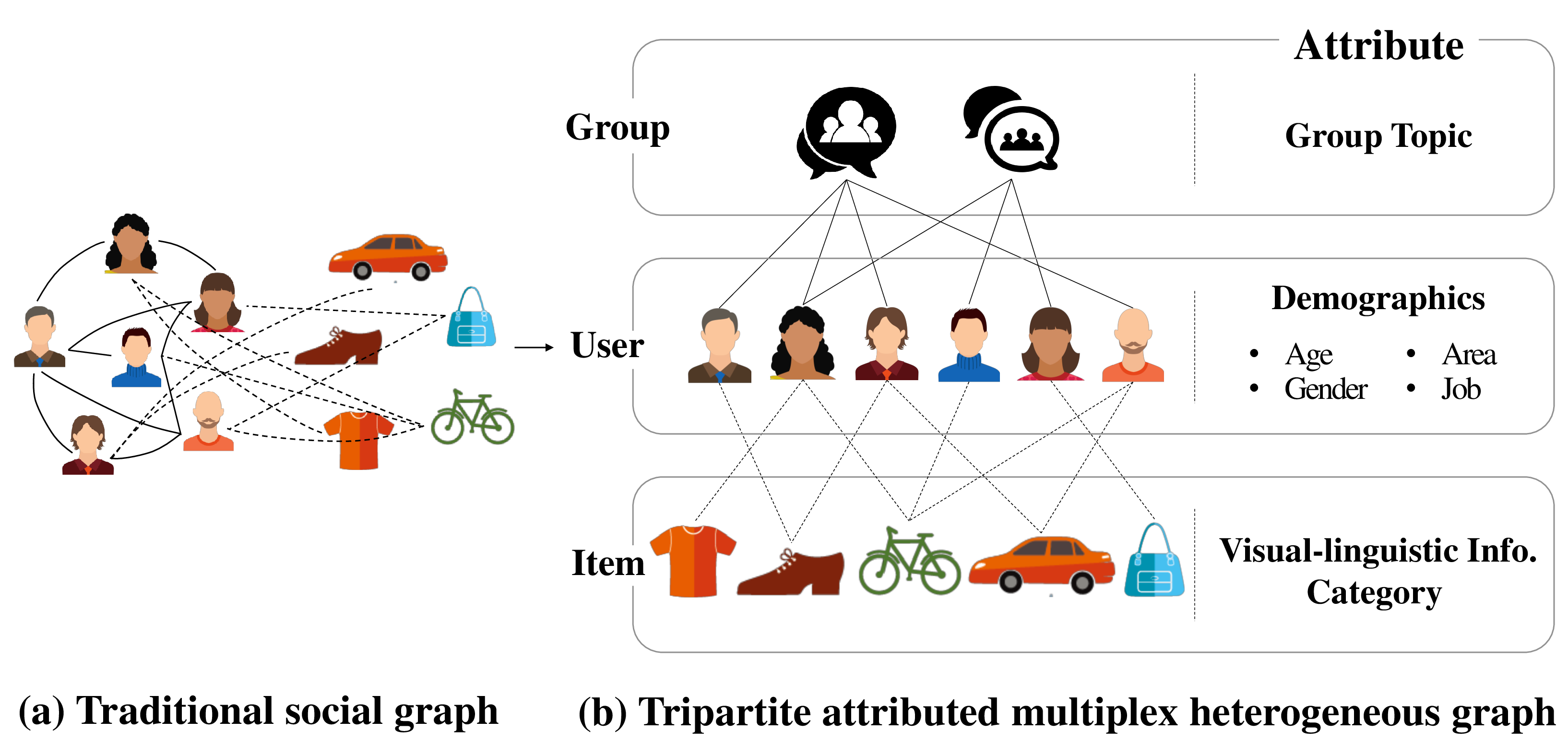}
  \caption{Social graph representation. (a) Traditional user-item graph for social recommendation. (b) Group-user-item tripartite attributed multiplex heterogeneous graph used in our task.}
  \label{fig:groupnode}
\end{figure}

In this paper, we propose a novel graph configuration and embedding method to tackle these four problems, called Heterogeneous Graph Propagation (HGP).
We introduce the concept of a group node that connects groups of related users defined by common social properties to mitigate the complexity of user connections.
A user node would belong to multiple group nodes, and there is no edge between user nodes. 
This configuration reduces computing time and memory while preserving social attributes and structures.
The group nodes have attributes, such as group topic, that represent the social properties.
Previous studies showed that exploiting different social relations can benefit the performance of social recommender system \cite{tang2012mTrust}.
We use these attributes as initial embedding of nodes.
This graph can be formulated as a tripartite attributed multiplex heterogeneous network \cite{cen2019representation} as illustrated in Figure \ref{fig:groupnode}.

To tackle the scalability issue, HGP sub-sample the nodes before propagating the graph, following the efficient sampling method \cite{chen2018fastgcn}.
Then, the HGP builds node embeddings separately for user-item graph and user-group graph.
To prevent the oversmoothing problem, it uses personalized PageRank scheme \cite{page1999pagerank,klicpera2018predict} when propagating node embedding through the whole graph structure.
The HGP handles the heterogeneity of graph in two ways; it applies different predicting functions for each node type and combines two types of the node embeddings from each graph with an attention-based function. 
Finally, the HGP recommends items to a user by retrieving the nearest items to the user in user-item joint embedding space.

We evaluate our HGP on a large-scale real-world dataset collected from a social application for our experiments.
The dataset includes 1.7M nodes consisting of 456K groups, 1.1M users, and 135K items. 
The nodes have multiple attributes such as group topic (group node), demographic properties (user node), visual-linguistic information, and category (item node).
The total number of edges is 4.7M. 
The experimental results show that our HGP outperforms competitive graph embedding methods.
Moreover, as the number of layers increases, the HGP can achieve better performance. It implies that propagating item preference of friends indeed help improve the performance of recommendation.

The main contributions of the paper are as follows:
\begin{itemize}
    \item We propose a novel graph-based recommendation method, Heterogeneous Graph Propagation (HGP), which prevents oversmoothing problem and handles heterogeneity of graph.
    \item We use a group-user-item tripartite attributed multiplex heterogeneous graph for social recommendation to reduce noise between user connections and the complexity of the graph.
    \item We show the effectiveness of our method on a large-scale, real-world dataset.
\end{itemize}

\section{Related Works}
There are many existing studies on the architecture of Graph Neural Networks (GNNs).
They can be categorized into five classes according to their structure \cite{zhonghan2019GNNsurvey}, i.e., graph auto-encoder, graph convolutional networks, graph attention networks, graph generative networks, graph spatial-temporal networks.
Also, \cite{cen2019representation} formulates a graph embedding task according to the graph configurations, e.g., node type (single or multiple), edge type (single or multiple), and attribute (use or non-use).
In GNNs terminology, we tackle attributed multiplex heterogeneous networks using a graph convolutional networks approach but with a personalized PageRank scheme \cite{klicpera2018predict, page1999pagerank}.

The social recommendation has attracted many researchers and practitioners with the popularity of social media \cite{irwin2010social}.
A typical assumption is that user preferences are similar to or influenced by socially connected friends, which is grounded on social correlation theory \cite{marsden1993social_influence, mcpherson2001homophily}.
Previous methods on social recommendation mainly have used collaborative filtering \cite{wang2018cfsocial} that decomposes user-user matrix and user-item matrix. 
There have been very few methods that apply deep learning or GNNs on social recommendation.
NSCR \cite{wang2017itemsilk} proposed a neural social collaborative ranking recommender system.
\cite{fan2019GNN_social, song2019GNN_session} presented GNNs architectures that deal with heterogeneity of social connections by calculating a weight value on each friend when aggregating neighborhoods.
However, the limitation is scalability. 
As many users are engaged in a social graph, connections become incredibly complex, and it may not be easy to calculate every weight values on all user combinations.
Moreover, neighborhood aggregation using attention mechanism is known to have high variance problem \cite{shchur2018pitfall}.
Thus, they only dealt with a small-sized social graph consisting of under 20K users \cite{fan2019GNN_social} or 141K users \cite{song2019GNN_session}, while our graph contains more than 1M users. 
Note that to solve heterogeneity of social connections, we use additional rich node attributes that are prevalent in real-world data.

\begin{figure*}[h]
  \centering
  \includegraphics[width=\linewidth]{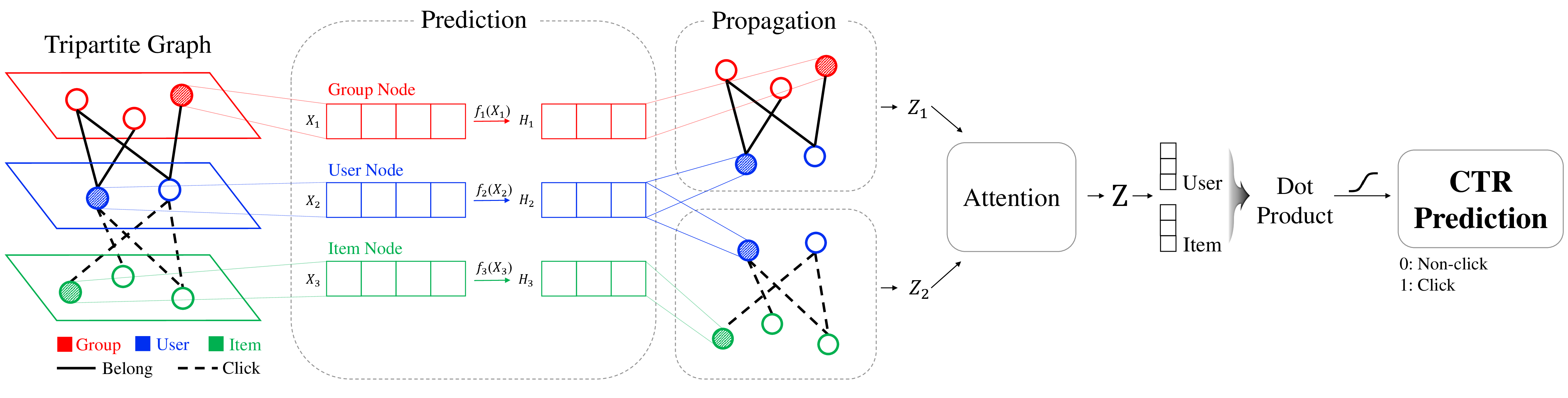}
  \caption{Schematic architecture of Heterogeneous Graph Propagation (HGP) as a social recommender system. HGP propagates neighborhoods independently for each edge type and then combines the final node representations with an attention model. We compute dot-product similarity between user and item representations to predict CTR.}
  \Description{.}
\end{figure*}

\section{Approximate personalized propagation of neural predictions}

Graph convolutional networks~\cite{kipf2017semi} and random walk~\cite{perozzi2014deepwalk} would cause oversmoothing if too many layers (or steps for the random walk) are used \cite{xu2018representation, klicpera2018predict}. 
Approximate personalized propagation of neural predictions (APPNP) \cite{klicpera2018predict} avoids the oversmooting by utilizing a propagation scheme derived from personalized PageRank \cite{page1999pagerank}. 
This algorithm adds a chance of teleporting back to the root node, balancing the needs of preserving locality and leveraging the information from a far neighborhood. 
The APPNP is efficient as it separates the neural network used for generating predictions from the propagation scheme. 

We first start with notations for homogeneous graph \begin{math}G=(V,E)\end{math} where \begin{math}V\end{math} and \begin{math}E\end{math} are nodes and edges respectively. The edges are described by the adjacency matrix \begin{math}A\in \mathbb{R}^{|V|\times |V|} \end{math}. To propagate self information of nodes to itself, the graph networks add self loops to the adjacency matrix: \begin{math}\Tilde{A}=A+I\end{math}. It is then symmetrically normalized as: \begin{math}\hat{A}=\Tilde{D}^{(-1/2)}\Tilde{A}\Tilde{D}^{(-1/2)}\end{math}, where \begin{math}\Tilde{D}\end{math} is the diagonal degree matrix of \begin{math}\Tilde{A}\end{math}. The nodes are initially represented by the feature matrix \begin{math}X\in \mathbb{R}^{|V|\times n} \end{math} where \begin{math}n\end{math} is the number of features. We compute \begin{math}X\end{math} by passing raw features of nodes into a node embedding network. The learnable parameters \begin{math}\theta\end{math} of APPNP only exist in the neural network \begin{math}f_\theta\end{math} that generates predictions \begin{math}H=f_\theta(X)\end{math}. With \begin{math}Z^{(0)}=H\end{math}, the propagation scheme at the $k$-th step is
\begin{equation}
\label{appnp_prop}
Z^{(k)} = (1-\alpha)\hat{A}Z^{(k-1)} + \alpha H,
\end{equation}
where \begin{math}\alpha\end{math} is the teleport probability of the personalized PageRank \cite{page1999pagerank}. The teleport probability \begin{math}\alpha\end{math} adjusts the effect of the neighborhood influencing each node. Note that there are no learnable parameters involving in the propagation scheme. This propagation scheme permits the use of far more propagation steps without leading to oversmoothing. The final node representation matrix \begin{math}Z^{(K)}\end{math} is then used for our tasks. 

\section{Heterogeneous Graph Propagation}

Items, users, and social relationships build a complex graph with multiple types of nodes and edges.
Naively applying the propagation scheme directly to the heterogeneous graph might inadvertently cause to bias training towards dominant edge types. 
To effectively handle different edge types, Heterogeneous Graph Propagation (HGP) propagates neighborhoods for each edge type independently and then combines the final node representations with an attention model. 
Also, HGP uses predicting neural networks separated for each node type considering heterogeneity of node attributes. 

\subsection{CTR Prediction}
In a heterogeneous graph $G$, there is a node type mapping function \begin{math}\phi:V\rightarrow O\end{math} and an edge type mapping function \begin{math}\psi:E\rightarrow R\end{math}. We denote \begin{math}A_r\end{math} as an adjacency matrix that only includes edges of type \begin{math}r\in R\end{math}. Following the similar notations from the previous section, we define a symmetrically normalized adjacency matrix \begin{math}\hat{A_r}\end{math} in the same way. 

We split the nodes by types: \begin{math}X_1, X_2, ..., X_{|O|}\end{math}, apply each predicting neural network: \begin{math}H_i = f_i(X_i)\end{math} and concatenate the results: \begin{math}H = [H_1, H_2, ..., H_{|O|}]\end{math}. Starting with \begin{math}Z^{(0)}_r=H\in\mathbb{R}^{|V|\times m}\end{math} for each edge type \begin{math}r\in R\end{math}, HGP uses similar  scheme as Equation \ref{appnp_prop}. The purpose of APPNP is not to learn deep node embedding, but to learn a transformation from attributes to class labels in the semi-supervised setting. HGP instead uses non-linear propagation with additional learnable weights to learn deep node representations:
 
\begin{equation}
Z^{(k+1)}_r = (1-\alpha)ReLU(\hat{A_r}Z^{(k)}_r W^{(k)}_H) + \alpha H.
\end{equation}


HGP combines the final node representation matrices with an attention model. Without loss of generality, we select \textit{i}-th node (row) from \begin{math}Z^{(K)}_{r}\end{math} for each edge type. We stack these vectors building a matrix \begin{math}Y_i\in\mathbb{R}^{|R|\times m}\end{math}. The attention model is a single layer Transformer \cite{vaswani2017attention}: 
\begin{equation}
Attention(Q, K, V) = softmax(QK^T/\sqrt{d_k})V,
\end{equation}
where \begin{math}d_k\end{math} is dimension of input queries and keys. Using this model, the HGP performs self attention to \begin{math}Y_i\end{math}:
\begin{equation}
Y'_i = Attention(Y_iW_Q, Y_iW_K, Y_iW_V),
\end{equation}
where query, key and value are same, except that different weight matrices are multiplied. Then, the HGP concatenates all rows of \begin{math}Y'_i\end{math} and pass it to a linear layer, generating a representation vector \begin{math}z_i\end{math} for \textit{i}-th node. 

In our application of social recommender system, we compute dot-product similarity between the user and item representations to predict CTR, i.e., click or not:
\begin{equation}
p_{i,j} = sigmoid(z^T_iz_j + x^T_ix_j),
\end{equation}
where \begin{math}x_i\end{math} is \textit{i}-th row vector of the feature matrix \begin{math}X\end{math}.
The ground truth of CTR is the existence of the edge connecting user and item nodes.
We optimize the model by reducing the cross-entropy loss with stochastic gradient descent algorithms.

\subsection{Sampling Strategy for Large-scale Heterogeneous Graph}
The recursive neighborhood expansion across layers needs massive time and memory to train with large and dense graphs.  
The node sampling methods, such as GraphSAGE\cite{hamilton2017inductive} and FastGCN \cite{chen2018fastgcn}, are generally adopted to overcome this problem. 
However, these methods are not suitable for heterogeneous graphs when there are dominant node types.
We handle this issue by adjusting the sampling probability to be proportional to the number of nodes for each type.
To reduce approximation variance, the sampling probability is also proportional to the degree of node \cite{chen2018fastgcn}. 
Additionally, we can also take advantage of inductive learning by adopting the sampling method. 

\section{Experiments}

\subsection{Datasets}

\begin{table}
  \caption{Statistics of the datasets.}
  \label{tab:stat}
  \begin{tabular}{cc}
    \toprule
    Node or edge types&Numbers\\
    \midrule
    \textit{User}  & 1,105,921\\
    \textit{Group} & 456,483\\
    \textit{Item} & 82,875\\
    \textit{Item}-\textit{User} & 3,746,650\\
    \textit{Group}-\textit{User} & 964,548\\
  \bottomrule
\end{tabular}
\end{table}

\begin{table}
  \caption{Hyperparameters of HGP.}
  \label{tab:hyper}
  \begin{tabular}{cc}
    \toprule
    Hyperparameters&Values\\
    \midrule
    Batch size & 1024\\
    Teleport probability $(\alpha)$ &  0.1\\
    \# of columns in $X$ $(n)$ &  16\\
    \# of columns in $H$ $(m)$ &  16\\
    \# of propagation $(K)$ &  10 \\
    Initial learning rate & $3e^{-7}$\\
    $\beta_1$ of Adam & 0.9\\
    $\beta_2$ of Adam & 0.999\\
    Dimension of $W_Q, W_K$ & ($m$, 16)\\
    Dimension of $ W_V$ & ($m$, 8)\\
    Dimension of $ W_H$ & ($m$, $m$)\\
    Dimension of node embedding $z_i$ & 16\\
    \ Avg. \# of sampled nodes & 10240\\
  \bottomrule
\end{tabular}
\end{table}

We use a dataset collected from a large-scale social network service. 
\textit{Group}, \textit{User} and \textit{Item} are three node types in the dataset. 
The group and user nodes are connected if the user belongs to the group.
The group nodes effectively reduce the number of edges and the complexity of paths compared to fully connecting all users in the group. 
The item and user nodes are connected when the user positively interacted with the item. 
There are 1,645,279  nodes connected with 4,711,208 edges. 
Table \ref{tab:stat} shows the overall statistics of the dataset. 
To enhance accuracy and generality, the nodes contain various attributes such as group topic of group node, demographic properties of user node, and visual-linguistic information and category of item node. 
These attributes are essential when predicting unseen nodes in test environments. 
We extract high-level features with BERT \cite{devlin2018bert} and VGG16 \cite{simonyan15} for visual-linguistic attributes. 
The BERT features (768-D) come from the last layer of [CLS] token, and the VGG16 features (4096-D) come from the FC6 layer.
We transform categorical attributes into dense features with linear embedding layers.
Finally, we aggregate all features to represent the nodes.
We use the first eleven days as a training set, the subsequent two days as a validation set, and the last four days as a test set.

\subsection{Comparable Models}

We compare our model with several models proposed for graph structures as well as a traditional model, i.e., Factorization Machine (FM).

\textbf{metapath2vec} \cite{dong2017metapath2vec}: 
It performs meta-path based random walk and leverages heterogeneous skip-gram model for node embedding. 
It only uses the identification information of nodes and does not cover attributes.
In our datasets, we choose \textit{G-U-I-U-G} as a meta-path which considers all node types.

\textbf{metapath2vec+EGES}: 
We modify metapath2vec to use the attributes. 
Following the attribute embedding methods from EGES \cite{wang2018billion}, it densely embeds each attribute and aggregates them by applying attention mechanism.


\textbf{MNE+EGES}: 
The nodes in MNE \cite{zhang2018scalable} use its distinctive embedding and several additional embeddings for each edge type, which are jointly learned by a unified graph embedding model. 
It conducts random walk for each edge type to generate sequences of nodes and then performs skip-gram algorithm.
The attribute embedding method is same as EGES.

\textbf{FastGCN} \cite{chen2018fastgcn}: 
The FastGCN is a homogeneous graph embedding method that directly subsamples the nodes for each layer altogether.
It is scalable but does not consider edge types. 
In our task, it uses the same attribute embedding method as HGP for initial node features.
	
\textbf{HGP}: The proposed model applies the personalized PageRank scheme and then combines node embeddings from the group-user graph and user-item graph with the attention model. 
It concatenates the attribute features and passes it to a single-layer perceptron for each node, generating the feature matrix \begin{math}X\end{math}. 
The predicting neural networks \begin{math}f_1,f_2,...f_{|O|}\end{math} are also single-layer perceptrons. 
The activation function of the single-layer perceptrons is ReLU.
We use Adam \cite{kingma2014adam} to minimize the cross-entropy loss of the predictions.
Table \ref{tab:hyper} summarizes other important hyperparameters. 

Since metapath2vec, metapath2vec+EGES, and MNE+EGES are unsupervised learning model, we train additional single-layer perceptron to predict CTR.
We measure the performance in terms of ROC-AUC, PR-AUC, and F1 score.
All implementations and the experiments were performed on NAVER SMART Machine Learning platform (NSML)~\cite{kim2018nsml, sung2017nsml}.

\subsection{Results}



\begin{table}[!t]
  \caption{Performance comparison of competing models. The hyphen `-' implies that we can not measure the stable performance due to the high variance of the results.}
  \label{tab:result}
  \begin{tabular}{cccc}
    \toprule
    Models&ROC-AUC&PR-AUC&F1\\
    \midrule
    FM & 0.5725  & 0.5654 & 0.5400 \\
    metapath2vec & 0.5  & - & -\\
    metapath2vec+EGES & 0.6136  & 0.6290 & 0.5604\\
    MNE+EGES & 0.6158  & 0.6307 & 0.5660\\
    FastGCN  & 0.6010  & 0.5937 & 0.5417 \\
    HGP (ours) & \textbf{0.6365}  & \textbf{0.6378} & \textbf{0.5967}\\
  \bottomrule
\end{tabular}
\end{table}

\begin{table}
  \caption{Learning time of our model according to the use of sampling method.}
  \label{tab:time}
  \begin{tabular}{cc}
    \toprule
    Method&Learning Time Per Epoch\\
    \midrule
    HGP w/o sampling & $\approx$ 45 hrs \\
    HGP  & $\approx$ 1.1 hrs \\
  \bottomrule
\end{tabular}
\end{table}

We report the experimental results of the competitors on Table \ref{tab:result}.
We found that the values of PR-AUC and F1-score are proportional to that of ROC-AUC.
The metapath2vec that does not use node attributes fails to learn CTR prediction.
The HGP outperforms the FastGCN, which is not suitable for heterogeneous graphs and suffers from the oversmoothing problem.
It also outperforms other recent heterogeneous graph models (metapath2vec+EGES and MNE+EGES).
Moreover, the validation loss of our model converges within a half day, which is suitable for the daily update of service model, required for industrial recommender systems.
In Table \ref{tab:time}, we compare the learning time of HGP according to the use of sampling scheme .

\begin{figure}[!t]
  \centering
  \includegraphics[width=\linewidth]{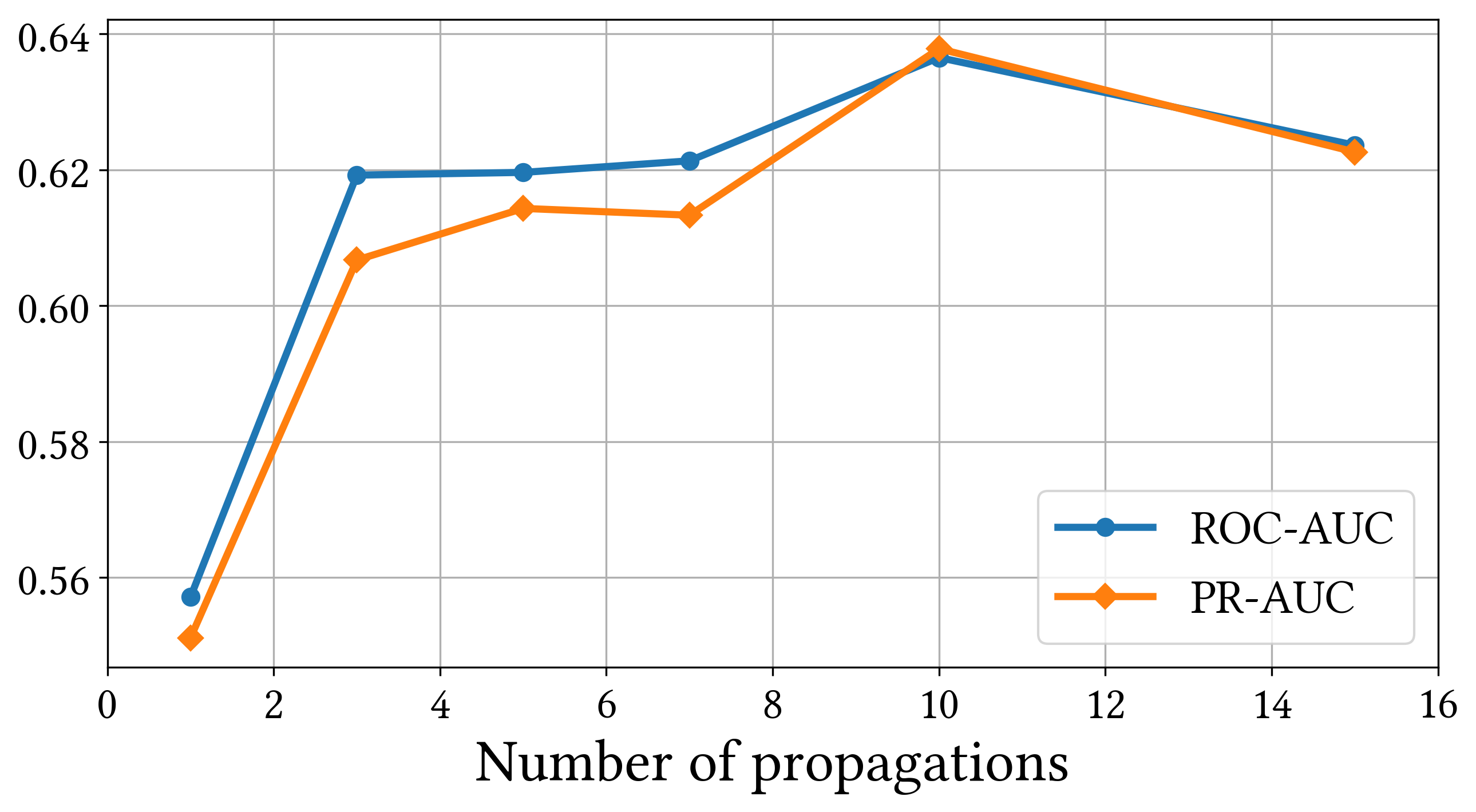}
  \caption{Performance comparison of HGP with different propagation steps.}
  \label{fig:prop}
\end{figure}

Figure \ref{fig:prop} shows the performance comparison of HGP with different propagation steps.
In our graph, HGP needs at least two propagation steps to know other members in a group (user $\rightarrow$ group $\rightarrow$ user).
If the number of propagation steps is three, it can approach other preferred items of users who have common preferences (user $\rightarrow$ item $\rightarrow$ user $\rightarrow$ item).
Considering the social correlation theory \cite{marsden1993social_influence, mcpherson2001homophily}, we can understand why the HGP with $k$=3 has better performance than that with $k$=1.
The performance of previous GCN architecture degrades as the number of propagation steps $k$ increases, even when the $k$ is two or three.
Overall, the HGP achieves the best performance at $k$=10 and successfully avoids the oversmoothing problem.


\section{Conclusion}
In this paper, we proposed a graph configuration, group-user-item tripartite attributed multiplex heterogeneous networks, for a social recommender system.
Our graph configuration reduces computing time and memory as the square of the number of nodes.
The attributes of group node help to exploit social relationship between users.
Additionally, we presented a graph-based recommendation method named Heterogeneous Graph Propagation (HGP).
To avoid the oversmoothing problem, the HGP propagates neighborhood information using the personalized PageRank scheme. 
The HGP can effectively handle the heterogeneity of a graph in two ways: 
1) It builds node embeddings separately for each edge type and combines them with the attention function.
2) It uses different predicting functions on each node type.
To handle the scalability issue, we adopted the sampling method suitable for the heterogeneous setting.
It is unable to train the dataset without the sampling due to lack of computing resources. 
We tested our model on the large-scale real-world dataset and showed that the HGP outperforms competitive graph embedding methods in terms of various metrics.
We plan to extend our graph by adding other social properties such as address, educational background, and common interests that would be effectively utilized for social recommender systems.


\begin{acks}
The authors appreciate Andy Ko for insightful comments and discussion.
\end{acks}

\bibliographystyle{ACM-Reference-Format}
\bibliography{sample-base}


\end{document}